\documentclass[12pt]{article}
\usepackage{graphics, color}
\usepackage{graphicx}
\usepackage{amssymb}
\pdfoutput=1

\newcommand{\sect}[1]{\section{#1}\setcounter{equation}{0}}

\def\gsim{\, \rlap{$>$}{\lower 1.1ex\hbox{$\sim$}}\,}
\def\lsim{\, \rlap{$<$}{\lower 1.1ex\hbox{$\sim$}}\,}

\begin{document}


\begin{titlepage}
\bigskip
\bigskip\bigskip\bigskip
\centerline{\Large Introduction to Gauge/Gravity Duality}
\medskip
\centerline{\large Lectures at TASI, June 1-7, 2010}
\bigskip\bigskip\bigskip
\bigskip\bigskip\bigskip

 \centerline{ {\bf Joseph Polchinski}\footnote{\tt joep@kitp.ucsb.edu}}
\medskip
\centerline{\em Kavli Institute for Theoretical Physics}
\centerline{\em University of California}
\centerline{\em Santa Barbara, CA 93106-4030}\bigskip
\bigskip
\bigskip\bigskip


\begin{abstract}
These lectures are an introduction to gauge/gravity duality, presented at TASI 2010.  The first three sections present the basics, focusing on $AdS_5 \times S^5$.  The last section surveys a variety of ways to generate duals of reduced symmetry.
\end{abstract}
\end{titlepage}
\baselineskip = 17pt
\setcounter{footnote}{0}

\tableofcontents

\sect{Generalities}

\subsection{The greatest equation}

A few years back, Physics World magazine had a reader poll to determine the Greatest Equation Ever, and came up with a two-way tie between Maxwell's equations
\begin{equation}
d*F=j\,,\quad dF=0\,,
\end{equation}             
and Euler's equation
\begin{equation}
e^{i\pi} + 1 = 0 \,.
\end{equation}             
The remarkable appeal of Euler's equation is that it contains in such a compact form the five most important numbers, $0, 1, i, \pi, e$, and the three basic operations, $+, \times,$ \^{}.  But my own choice would have been Maldacena's equation
\begin{equation}
{\rm AdS} = {\rm CFT} \,,
\end{equation}
because this contains all the central concepts of fundamental physics: Maxwell's equations, to start with, and their non-Abelian extension, plus the Dirac and Klein-Gordon equations, quantum mechanics, quantum field theory and general relativity.  Moreover, in addition to these known principles of nature, it contains several more that theorists have found appealing: supersymmetry, string theory, and extra dimensions, and it ties these all together in an irreducible way.  Also, while Euler's equation is a bit of an oddity, the relation AdS = CFT is just the tip of a large iceberg, it can be deformed into a much large set of gauge/gravity dualities.

So I get to teach you about this, trying to focus on things that you will need in the many upcoming lectures which will use the left-hand side of the equation, string theory and gravity, to learn more about the the right-had side, quantum field theory.  Of course it's also interesting to use the equation in the other direction, and maybe some of that will sneak in.  Roughly speaking, today's lecture will be a conceptual overview, lecture 2 will give some essential details about the two sides of the duality, lectures 3 and 4 will work out the dictionary between the two sides focussing on the familiar $AdS_5 \times S^5$ example, and lecture 5 will discuss generalizations in many directions.

Let me start by noting a few other reviews.  The early MAGOO review~\cite{Aharony:1999ti} contains a detailed summary of the early literature, in which many of the basic ideas are worked out.  The 2001 TASI lectures by d'Hoker and Freedman~\cite{D'Hoker:2002aw} are thorough and detailed, particularly with regard to the constraints from supersymmetry and the conformal algebra, and the calculation of correlation functions.  McGreevy's course notes~\cite{mcgnotes} are similar in approach to my lectures.

\subsection{A hand-waving derivation}

I am going to first motivate the duality in a somewhat unconventional way, but I like it because it connects the two sides, gauge theory and gravity, without going directly through string theory (as do many of the applications), and it allows us to introduce many ideas that will be important later on.  So let me start with the question, is it possible to make the spin-2 graviton as a bound state of two spin-1 gauge bosons?  With the benefit of generous hindsight, we are going to make this idea work.  To start off, there is a powerful no-go theorem that actually forbids it~\cite{Weinberg:1980kq}.  Theories without gravity have more observables than theories with gravity (local operators, in particular, since there is no invariant local way to specify the position in general relativity), and this leads to a contradiction.  Specifically, Weinberg and Witten show that if there is a massless spin-2 particle in the spectrum, then the matrix element
\begin{equation}
\langle {\rm massless\ spin\ 2}, k | T_{\mu\nu} |  {\rm massless\ spin\ 2}, k' \rangle 
\end{equation}
 of the energy momentum tensor (which exists as a local observable in the gauge theory) has impossible properties.

Of course, to prove a no-go theorem one must make assumptions about the framework, and it often proves to be the case that there is some assumption that is so natural that one doesn't even think about it, but which turns out to be the weak link.  The Coleman-Mandula theorem, classifying all possible symmetries of the S-matrix~\cite{Coleman:1967ad}, is a classic example.  This paper played an important role in its time, ruling out a class of ideas in which spin and flavor were unified in $SU(6)$.  However, it made the unstated assumption that the symmetry generators had to be bosonic,\footnote{Rereading Ref.~\cite{Coleman:1967ad}, this assumption seems to have made its entrance at the point where the symmetry generators are diagonalized, which can't be done for a nilpotent operator.} which was sufficient for the immediate purposes but missed the possibility of supersymmetry.  The more powerful a no-go theorem, the deeper its counterexamples.\footnote{Perhaps this is what Bohr meant when he said ``It is the hallmark of any deep truth that its negation is also a deep truth.''}

The reason for going through this is that the no-go theorem is indeed wrong, but to violate it we have to recognize a deep property of quantum gravity, the holographic principle~\cite{'tHooft:1993gx,Susskind:1994vu}.  The entropy of a black hole is proportional to its area in Planck units, and this is the largest possible entropy for a system with given surface area.  This suggests that quantum gravity in any volume is naturally formulated in terms of degrees of freedom on its surface, one per Planck area.  Thus we see the hidden assumption, that the graviton bound state moves in the same spacetime as its gauge boson constituents; rather, it should move in one additional dimension.  Of course, there might be other ways to violate the theorem that will turn up in the future.

So how is the two-gauge boson state in four dimensions supposed to correspond to a graviton in five dimensions?  With the benefit of hindsight, there are several places in QCD phenomenology where the size of a gluon dipole, the magnitude $z$ of the separation, behaves like a spacetime coordinate.  In color transparency~\cite{Bertsch:1981py}, and in the BFKL analysis of Regge scattering~\cite{Lipatov:1985uk}, interactions are approximately local in $z$ and the pair wavefunction satisfies a five-dimensional wave equation.  So when you look at the gluon pair you picture it as a graviton four of whose coordinates are the center of mass of the pair, and the fifth is the separation.

We need just two more ingredients to make this idea work, but first we will make an excursion and discuss the shape of the five dimensional spacetime.  Quantum field theory is nicest when it applies over a wide range of scales, so it is natural to consider scale-invariant theories first.  If we rescale the system, the c.m.\ coordinates $x$ and the separations $z$ scale together.  The most general metric respecting this symmetry and the symmetries of the four-dimensional spacetime is
\begin{equation}
ds^2 = \frac{L^2 dz^2 +  L'^2 \eta_{\mu\nu} dx^\mu dx^\nu}{z^2} \to L^2 \frac{dz^2 +  \eta_{\mu\nu} dx^\mu dx^\nu}{z^2} \,,
\label{adsmet}
\end{equation}  
where I have rescaled $z$ in the last form so as to emphasize that there is only one scale $R$.

This is the metric of anti-de Sitter space, in Poincar\'e coordinates.  If we replace the $z$ with $t$ in the denominator we get de Sitter space, the approximate geometry of our own accelerating spacetime.  Certainly one of the major frontiers in gauge/gravity duality, though not one that I will focus on, is to figure out how to interchange $z$ and $t$ in the dual field theory: it appears to require great new concepts.  Anti-de Sitter spacetime is not expanding but warped: clocks that run at the same rate in inertial coordinates run at different rates in terms of $x^0$, depending on where they are in $z$. 

To finish off our `derivation' of the duality, we need two more ingredients.  The first is a large number of fields.  We want the AdS scale $L$ to be large compared to the Planck length $L_{\rm P}$, so that we can use Einstein gravity.  This means that we can fit a large black hole into the space, one with many Planckian pixels and so a large entropy, and so the field theory had better have a correspondingly large number of degreees of freedom.  As we will discuss in more detail later, the number of fields is a power of $L/L_{\rm P}$, depending on the example.  The other ingredient we need is strong coupling, so that the gauge boson pair behaves like a graviton and not like a pair of gauge bosons.\footnote{Of course, it will then mix with states having more constituents, but one can still retain a bit of the basic idea that the graviton spin comes from two `valence' gauge bosons.}  I should say VERY strong coupling, much larger than one, to get a limit in which the gravitational description is quantitative.  

Thus we have two necessary conditions for the duality, many fields and very strong coupling.  The second condition can actually be made a bit stricter, as we will see: we need that most operators get parametrically large anomalous dimensions.  In this form, these necessary conditions are actually likely to be sufficient, as shown in part in Ref.~\cite{Heemskerk:2009pn}.  Large anomalous dimensions clearly require very strong coupling, but do not necessarily follow from it as we will see in an example later.

These conditions of many fields and strong coupling will reappear at various points in these lectures.  Clearly they play a controlling role in the applications, as to whether we have a quantitive description or merely a `spherical cuprate,' a solvable model that captures the qualitative physics that we wish to understand.  For example, in the application to heavy ion physics, the coupling seems to be of order one, midway between weak and strong.  The strong coupling picture is not quantitative in a precise way, but seems to do better than perturbation theory on many qualitative properties.  Naively the number of fields is $N^2 - 1 = 8$ for the gluon states, which is a modestly large number, though it has been noted that the parameter $N/N_{\rm flavor}$ is only unity, and there should be corrections of this order.  Fortunately the large-$N$ approximation seems to be fairly robust even at small values.  

In the condensed matter applications, again the relevant couplings are of order one at a nontrivial fixed point, where a strong coupling expansion has a chance to capture things that a weak coupling expansion cannot.  There is no large $N$, but condensed matter theorists in the past have not been above introducing a large-$N$ vector index in order to get a tractable system.  The large-$N$ vector is a mean-field approximation; this is true for the large-$N$ matrix limit as well, though in a more subtle and perhaps more flexible way~\cite{Witten:1979pi}: expectations of products of color singlets factorize, but there is a very large number of color singlet fields.

Notice that I have not yet mentioned supersymmetry.  However, it tends to enter necessarily, through the requirement of very strong coupling.  Quantum field theories tend to become unstable at strong coupling, through the production of pairs whose negative potential energy exceeds their kinetic energy.  In continuum theories this can happen at all scales, and the theory ceases to exist.  Supersymmetry protects against this: schematically the Hamiltonian is the sum of the squares of  Hermitian supercharges, $H = \sum_i Q_i^2$, so the energy is bounded below.  Thus, we can break the supersymmetry softly and still have a duality, but if we break it at high energy we lose the theory.  We may have to be satisfied with metastability, but that is OK, we probably live with that in our own vacuum.

We have also not mentioned strings, we seem to have found a theory of quantum gravity that uses only gauge theory as a starting point, but in a range of parameters for $N$ and coupling that is not so familiar.  But of course what happens is that when we get gravity in this way we get everything else as well, the strings, branes, extra dimensions and so on.  Again this was anticipated by 't Hooft~\cite{'tHooft:1973jz}, who argued that the planar structure of large-$N$ gauge theory made it equivalent to a theory of strings.

\subsection{A braney derivation, for $AdS_5 \times S^5$}

Let me now outline, much more briefly, the original argument for the duality~\cite{Maldacena:1997re}.\footnote{Note also Ref.~\cite{Polyakov:1997tj}, which reaches a similar picture by very different reasoning.}  This will have the advantage that it gives specific constructions of gauge/gravity duals.  Consider a stack of $N$ coincident D3-branes, in the IIB superstring.  In string perturbation theory, each additional world-sheet boundary on the branes brings in a factor of  the string coupling $g = e^{\Phi}$ from the genus and a factor of $N$ from the Chan-Paton trace.  Thus perturbation theory is good when $gN\ll 1$ and breaks down when $gN \gg 1$.\footnote{Of course the same is true for other D$p$-branes, but the rest of the story is simplest for $p = 3$.  We will consider other cases later.}  Now, there are black branes~\cite{Horowitz:1991cd} that source the same Ramond-Ramond fluxes as D-branes~\cite{Polchinski:1995mt}.  The black 3-brane metric for $N$ units of flux (in string frame where the physical string tension is $\alpha'$) is
\begin{eqnarray}
ds^2 &=& H^{-1/2}(r) \eta_{\mu\nu} dx^\mu dx^\nu + H^{1/2}(r) dx^m dx^m \, ,\quad
\mu,\nu \in 0,\ldots,3\,,\quad m,n \in 4,\ldots,9 \,,
 \nonumber\\
H &=& 1 + \frac{L^4}{r^4} \,,\quad L^4 = 4\pi g N \alpha'^2 \,,\quad r^2 = x^m x^m \,.
\label{black}
\end{eqnarray}
You will have to take my word for this solution here, but we will make a useful check later.

The harmonic function $H$ is singular as $r \to 0$, but the metric has a nice behavior in the `near-horizon' small $r$ limit,
\begin{equation}
ds^2 \to \frac{r^2}{L^2} \eta_{\mu\nu} dx^\mu dx^\nu + \frac{L^2}{r^2} dr^2 + L^2 d\Omega^2_{S^5}
\,. \label{ads5s5met}
\end{equation}
The first two terms are again the $AdS_5$ metric, with the coordinate $r = R^2/z$, while the last term is the metric of $S^5$.  The curvature radius is $L$ for both factors.  When $gN \gg 1$ this length is large in string units, so the low energy supergravity theory is a good effective description, but it breaks down when $gN \ll 1$.  This is just complementary to the perturbative description: the system looks like D-branes in one regime and black branes in another.  By varying $g$ one can move adiabatically between these.  This was the strategy used to count the states of a black hole~\cite{Strominger:1996sh}, though in a different brane system that has a finite zero-temperature entropy, as we will discuss later.

This is not yet a duality; we have a single theory, the IIB string theory, describing both limits, but in different approximations: perturbation theory around flat spacetime with branes at small $gN$, and perturbation theory around the curved black brane spacetime (with no branes and no open strings) at large $gN$.  To find a duality let us look at the low energy limit of both descriptions.  In the small-$gN$ description, this consists of the massless open and closed strings.  The massless open strings, ending on the D3-branes, are the usual $U(N)$ adjoint gauge field and collective coordinates, as well as their fermionic partners, while the massless closed strings are the supergravity multiplet.  The open strings remain interacting at low energy, because the gauge coupling is dimensionless in 3+1 dimensions, but the closed strings have irrelevant interactions and decouple.  In the large-$gN$ description, we again have the massless closed strings away from the brane, but there are also states whose energy is small because they are in the $AdS_5 \times S^5$ region at small $r$, where the warp factor $g_{00}$ is going to zero; these include not just the massless states, but any massive string state will have an arbitrarily small energy as $r \to 0$.  

Let us ignore the massless closed strings away from the brane, which are decoupling in both pictures.  Concretely, consider the scaling $r \to r/\zeta$, $x^\mu \to \zeta x^\mu$.  The only effect on the black brane metric~(\ref{black}) is that the 1 in the harmonic function becomes $\zeta^{-4}$, and scales away as $\zeta\to\infty$ leaving $AdS_5 \times S^5$.  In the D3-brane picture this is a symmetry of the low energy gauge theory, while the closed string phase space volume scales to zero, and massive open string effects are suppressed by powers of $\alpha'/x^2 \sim \zeta^{-2}$.  In Ref.~\cite{Maldacena:1997re} this limit is described in terms of scaling $\alpha'$ to zero while holding $x^\mu$ fixed, with appropriate scalings elsewhere; I have always found it harder to think this way, but it is the same in terms of dimensionless ratios.

Now, if we make the innocuous-sounding assumption that taking the low energy limit commutes with the adiabatic continuation in $g$ we get a remarkable result.  At weak coupling we have the gauge theory, and at strong coupling we have all the string states in the $AdS_5 \times S^5$ region.  The assumption that the limits commute means that the strongly coupled gauge theory is identical to the full string theory in $AdS_5 \times S^5$.  This is now a duality, the statement that two seemingly different theories describe the same system, but in different limits.

This `derivation' of the duality seems rather slick, but it explained why very different calculations in the two pictures were giving identical answers, and has been supported by many further checks over time.  How could the argument fail?  The most obvious issue is that the gauge theory could have a phase transition as we vary from weak to strong coupling, but the supersymmetry strongly restricts this possibility.  In particular, the property $H = \sum_i Q_i^2$ implies that any supersymmetric vacuum is a minimum of the energy, zero, so we can't have some other vacuum cross to lower energy at some intermediate coupling.\footnote{This would become an issue however if we try to extend the argument to nonsupersymmetric configurations, as we might want to in condensed matter systems.  I will discuss an examples in Sec.~4.9.}  
We can imagine more intricate ways that the duality might fail, and I will say a bit more about this later, but it is very hard to think of an alternative that is consistent with all the evidence, and the simplest conclusion is that the duality, remarkable though it be, is  true.

There are many other weak-strong duality conjectures.  Some relate one field theory to another, and some relate one string theory to another or to M theory, but here we have a duality between a quantum field theory and a string theory.  From the point of view of trying to construct a theory of quantum gravity this duality is particularly striking, because it allows us to reduce it to the problem of constructing a QFT, solved by Wilson; of course there is much more to say on this point, but it is not the focus of these lectures.  What most of these dualities have in common is that there is no explicit derivation, but rather there is the kind of plausibility argument we have just made, combined with many tests giving circumstantial evidence.  Some QFT-QFT dualities can be derived explicitly, like bosonization and Ising self-duality in 1+1 dimensions, and Abelian dualities in higher dimensions.  These seem as though they should be prototypes for all the other dualities, but somehow the direct steps used to derive them are not enough, and some big new idea is needed.

\subsection{Statement of the duality}

The precise statement is that 
\begin{equation}
D=4,\  {\cal N} = 4,\ SU(N)\ \mbox{Yang-Mills} \ =\ {\rm IIB\ string\ theory\ on}\ AdS_5 \times S^5 \,.
\end{equation}
We will give more detail about each side in the next lecture.  Again, this is just one example of a very large number of such dualities.  We focus on it, as do most other introductions, because it is both the simplest example, and the one whose field theory side is the most familiar and relevant to many applications.  The equality means a one-to-one mapping of the spectra, at any given value of the energy and other quantum numbers.  It also includes equality of observables, namely the correlation functions of operators with an appropriate dictionary between the two sides.

A minor aside that may occur to the reader: we have replaced the $U(N)$ of the brane stack with $SU(N)$.  The missing $U(1)$ represents the supermultiplet containing the overall collective motion of the stack.  In the black brane picture, before we take the near-horizon limit, this is a fluctuation of the spacetime geometry.  The mode is peaked in the transition region between flat spacetime, where the 1 in $H(r)$ dominates, and the near-horizon region, where the $L^4/r^4$ dominates.  In the $AdS_5 \times S^5$ limit, this moves away, and the collective mode is trivial there.

Another aside: the holographic argument led to one extra dimension, but now we have five more from the $S^5$.  Still, the Poincar\'e and radial dimensions will play a central role, and for much of the discussion we can reduce on the $S^5$.  Note that the warping acts only on the Poincar\'e dimensions, the $S^5$ having a constant radius.

On the gauge side we have the two parameters $g^2_{\rm YM}$ and $N$.  On the string side these map to the string coupling $g$ and again $N$, where the latter is now interpreted as the number of units of five-form flux on the $S^5$, which counts the D3-branes in the stack.  The map between the couplings is standard for D3-branes, $g = g_{\rm YM}^2/4\pi$, with action normalized $-\frac{1}{2g_{\rm YM}^2}{\rm Tr}\, F_{\mu\nu} F^{\mu\nu}$.  

The parameters on the gravity side are also usefully expressed in terms of ratios of length scales.  We have already noted that
\begin{equation}
L/L_{\rm string} \equiv L/{\alpha'^{1/2}} = (4\pi g N)^{1/4} = (g_{\rm YM}^2N)^{1/4} = \lambda^{1/4}\,,
\label{lls}
\end{equation}
where in the last line we have introduced the 't Hooft parameter $\lambda = g_{\rm YM}^2N$.  The other relevant length is Planck's.  In general spacetime dimension $D$ it is natural to work with the reduced Planck length $\hat L_{{\rm P},D}$, such that the coefficient of the Ricci scalar in the action is $1/2\hat L_{{\rm P},D}^{D-2}$; in four dimensions $\hat L_{\rm P,4} = (8\pi)^{1/2} L_{\rm P}$ in terms of the usual Planck length.
Then in string theory $\hat L_{\rm P,10}^8 = \frac{1}{2} (2\pi)^7  g^2 \alpha'^4$, and 
\begin{equation}
L/\hat L_{\rm P,10} = 2^{-1/4} \pi^{-5/8} N^{1/4} \,.
\label{llp}
\end{equation}
A classical spacetime description requires the ratios~(\ref{lls}) and ({\ref{llp}) to be large, and so as anticipated in the handwaving argument we need the coupling $\lambda$ and the number of fields $N$ to be large.

The evidence for the duality takes many varied forms, some of which will be discussed as we go along:
\begin{enumerate}
\item The symmetries on the two sides match.
\item The spectra of supersymmetric states match.  This includes, for example, all modes of the graviton in $AdS_5 \times S^5$.
\item Amplitudes which are protected by supersymmetry and so can be compared between the two sides are equal.
\item When we perturb the duality in ways that break some of the supersymmetry and/or conformal symmetry, the geometry realizes the behaviors expected in the field theory, such as confinement.
\item There are higher symmetries on both sides, which allow some quantities to be calculated for {\it all} $g$, with apparent consistency.
\item Matching long string states can be identified on both sides.
\item The predictions of the duality for strongly coupled gauge theories can be compared with numerical calculations in those theories, using both light-cone diagonalization and lattice simulation (these are computationally challenging, but progressing).
\item The predictions agree with experiment!
\end{enumerate}
The next-to-last item, and especially the last, are currently at a crude level of accuracy.  Probably I have missed some good tests.  Essentially, every time one applies the duality one is testing it, because there is always the possibility of some absurd consequence.

It is sometimes asserted that the evidence supports only a weak form of the duality, but it is not clear what a sensible weak form would be.  Suggestions include only the supergravity states, but the number of states in the gauge theory is much larger than this, and some can be identified clearly with string states.  Another weak form would be to hold only in the extreme large-$N$ limit.  However, this is not consistent (unitary) by itself on either side of the duality, and unitarity largely determines the $1/N$ expansion on the QFT side and the gravitational loop expansion on the string side; it follows from (\ref{llp}) that these are expansions in equivalent parameters.  Yet another weak form would be to hold only perturbatively in $1/N$ and not exactly.  However, the most important nonperturbative phenomena is present on both sides: the integer property of $N$.  In the $1/N$ expansion $1/N$ is a continuous parameter, but in reality it is discrete with an accumulation point at zero.  The string side knows about this because the 5-form flux satisfies a Dirac quantization condition.  

Thus, by far the simplest way to account for all the facts is that the duality is an exact statement.  Of course, we only have an explicit construction of the theory on the QFT side,\footnote{When I say this, I am thinking of the lattice regulator, with supersymmetry broken but restored in the continuum limit.  For the 3+1 example that we are considering, a skeptic can still doubt whether this limit is controlled, because the coupling is strong, but we will see other examples where the theory flows from a superrenormalizable gauge theory and so the continuum limit is much simpler.} so I mean that the QFT must agree with all of the approximations we have to the string theory, and with any future constructions of the theory.  Anyway, the QFT is fully quantum mechanical and consistent, and as we have noted it includes all the graviton states (with the right trilinear interactions), so at the very least it is {\it some} theory of quantum gravity.

One should note that there are local symmetries on both sides of the duality, the $SU(N)$ gauge symmetry in the QFT and coordinate invariance and local supersymmetry on the AdS side.  These are different and neither contains the other.  It is an important general principle that dualities acts only on the physical quantities and not on the redundant variables that we use to construct them.

We conclude with some homework problems to think about before the next lecture:
\begin{enumerate}

\item[Ex.\ 1.]It is obviously absurd to claim that a four-dimensional quantum field theory is the same as a ten-dimensional string theory.  Give one or more reasons why it can't be true.

\item[Ex.\ 2.] Figure out why your answer to the previous problem is wrong.
\end{enumerate}

\sect{The two sides}

\subsection{The gravity side}

The massless sector of the IIB string theory, IIB supergravity, controls the physics on long distance scales on the gravity side.  The fields are the metric $G_{MN}$, the dilaton, and a 3-form Neveu-Schwarz (NS) field strength $H_{MNP}$, and 1-, 3- and self-dual 5-form Ramond-Ramond (RR) field strengths $F_M$, $F_{MNP}$, and $F_{MNPQR}$ (I'll use capital Roman indices for spacetime, saving Greek for the Poincar\'e directions).  You can find the action in your favorite textbook.  Of course the theory also includes excited string states, and D($-1$), D1, D3, D5, and D7-branes, as well as NS5-branes and the various $(p,q)$ bound states.  All these branes are much heavier than the massless fields, but they exist in the $AdS_5 \times S^5$ region and represent various excitations of the gauge theory.

We could write out the field equations here and verify that $AdS_5 \times S^5$ is a solution, but instead I will do something a bit cruder, which gives some useful insight.  Leaving out numerical constants, the relevant terms in the action are
\begin{equation}
S \sim {\alpha'^{-4}} \int d^{10}x \, \sqrt{-G} ( e^{-2\Phi} R - F_{MNPQR} F^{MNPQR} ) \,.
\end{equation}
The 5-form is self-dual, which complicates the action principle.  One way to deal with this, which is most efficient for the energetic argument that we are going to make, is to treat the fully spatial components as the independent fields.\footnote{In fact, even for forms other rank, we have the option of using either a $q$-form or a Hodge-dual $(D-q)$ form, and for energetics it is always useful to work with the one that is spatial, else there are surface terms that enter.  Note that any dilaton or moduli dependence in the field strength action gets inverted for the Hodge dual. \label{fnhodge}}  Now, we are going to make a Kaluza-Klein reduction on the $S^5$, which is taken to have radius $r$.  The Dirac quantization condition gives $F_{MNPQR} \sim N \alpha'^2$, and so 
\begin{equation}
S_5 \sim {\alpha'^{-4}} \int d^{5}x \, \sqrt{-G_5} r^5 ( e^{-2\Phi} R_5 + e^{-2\Phi} r^{-2} -  \alpha'^4 N^2 r^{-10}) \,.
\end{equation}
The second term is from the curvature of the sphere, and in the third we include $r$-dependence from the inverse metric.  To interpret this in terms of an effective potential, rescale $G_5 \to r^{-10/3} e^{-4\Phi/3} G_5'$ to get
\begin{equation}
S_5 \sim {\alpha'^{-4}} \int d^{5}x \, \sqrt{-G_5'} ( R'_5 + e^{4\Phi/3} r^{-16/3} -  \alpha'^4 N^2 e^{10 \Phi/3} r^{-40/3} ) \,.
\end{equation}
Then 
\begin{equation}
V(r,\Phi) \sim - \frac{1}{x^{4/3}} + \frac{\alpha'^4 N^2}{x^{10/3}}
\end{equation}
where $x = r^4 e^{-\Phi}$.  The negative term from the curvature of $S^5$ dominates at large $x$, while the positive flux term dominates at small $x$, leaving a minimum at $x \sim \alpha'^2 N$, or
\begin{equation}
r^4 \sim \alpha'^2 N e^{\Phi} \,.
\end{equation}
This is in agreement with the exact solution asserted before, both the scaling and the presence of a flat direction in the potential so that the dilaton is undetermined.  The minimum is at negative potential, giving rise to an $AdS_5$ solution.

This energetics argument is similar to the way one studies stabilization of moduli in string compactification~\cite{Silverstein:2004id}.  It is crude by comparison to the sophisticated methods that are employed to find anti-de Sitter solutions in supergravity, but it is a useful complement to these.  Some examples:

\begin{enumerate}

\item[Ex.\ 3.] M theory has just a metric and a four-form flux, and the only length scale is the Planck scale $L_{\rm M}$.  Use the potential method to find $AdS_7 \times S^4$ and $AdS_4 \times S^7$ solutions, and determine how the radii of the spheres scale with the number of flux units (note footnote~\ref{fnhodge}).

\item[Ex.\ 4.] In the IIB theory find an $AdS_3 \times S^3 \times T^4$ solution with $Q_5$ units of RR 3-form flux on the $S^3$ and (again using footnote~\ref{fnhodge}) $Q_1$ units of RR 7-form flux on $S^3 \times T^4$.  Identify flat directions, and the scalings of the radii with the number of units of each of the fluxes.

\end{enumerate}

All these solutions have another important property that is more easily seen in 10 dimensions than in the 5-dimensional reduction: the curvature radii of the spherical factors and the AdS factor are of the same order.  If one considers the Einstein equations along the AdS space and the sphere, one can conclude that these curvature terms are of the same order unless there are cancellations.  In fact, in all explicitly known examples the radii are similar.  A framework for constructing AdS/CFT duals with a large hierarchy is given in Ref.~\cite{Polchinski:2009ch}, but many details remain to be filled in.  Again, from study of moduli stabilization~\cite{Kachru:2003sx} we know that there are many string theory solutions where all the compact directions are much smaller than the AdS radius, but the CFT duals are not known.  Making progress in this direction is important for the nonperturbative construction of the landscape, and also for the top-down construction of applied dualities.

\subsection{More about anti-de Sitter space}

Anti-de Sitter space has unusual properties that play a key role in the duality.  Referring to the metric~(\ref{adsmet}), the radial coordinate range is $0 < z < \infty$, and the null geodesics are
$z = \pm (t - t_0)$.  These reach the boundary $z=0$ in finite time, and so boundary conditions must be imposed there in order to have a well-defined system.  These boundary conditions also provide a natural set of observables, which generally are complicated to define in quantum gravity.
On the other hand, the geodesics take infinite time to reach or emerge from $z = \infty$.

The gravitational reshift means that invariant time intervals and coordinate time intervals are related  as $d\tau = L dx^0/z$.  Correspondingly, the energy $\cal E$ of an excitation as seen by an invariant observer, and its Killing (conserved) energy $P_0$ conjugate to $x^0$ are related
\begin{equation}
P_0 = L {\cal E}/z \,.
\end{equation}
A given bulk excitation has an energy that increases when it moves nearer the boundary, which is another way of thinking about the origin of $z$ in the CFT.  For Kaluza-Klein excitations near rest ${\cal E} \sim 1/L$ and so $P_0 \sim 1/z$.  It is closely related to the earlier interpretation in terms of gluon separation: if we work in terms of the internal momenta for the gluon we have $p_z \sim 1/z \sim P_0$.\footnote{Of course there are also bulk excitations associated with other energy scales, like the string scale, which would introduce a possibly large dimensionless constant into the relation $P_0 \sim 1/z$.  In this case it seems more general to associate $z$ with the size of the state~\cite {Susskind:1998dq}.}

The Poincar\'e coordinates~(\ref{adsmet}, \ref{ads5s5met}) are the natural ones for applications, but this space is not geodesically complete: the $z = \infty$ limit is a horizon, and a bulk excitation can actually fall through it.  A nice way to describe the global space $AdS_{D+1}$ is by embedding it in $D+2$ dimensions with signature $(D,2)$,\footnote{I will always use $D$ for the spacetime dimension of the field theory and $D+1$ for the dimension of the AdS space.}
\begin{equation}
ds^2 =  L^2 \gamma_{ab} dX^a dX^b\,, \quad \gamma_{ii} = +1\,,\ i = 1, \ldots, D \,, \quad\gamma_{00} = \gamma_{D+1,D+1} = -1 \,.
\end{equation}
The embedding is
\begin{equation}
\gamma_{ab} X^a X^b = - 1 \,.
\end{equation}
Defining $U = X^D - X^{D+1}$ and $V = X^D + X^{D+1}$, this is
\begin{equation}
ds^2 = L^2(\eta_{\mu\nu} dX^\mu dX^\nu + dU dV) \,,  \quad \eta_{\mu\nu} X^\mu X^\nu + U V = -1 \,,
\end{equation}
where $\mu$ runs over the Poincar\'e directions.
Solving for $V$ and defining $x^\mu = X^\mu/U$, $z = 1/U$, the metric in the region $U > 0$ takes the Poincar\'e form~(\ref{adsmet}).  The extension to negative $U$ doubles this space, and we can go further: the sum $(X^0)^2 + (X^{D+1})^2$ is positive definite, so there is a noncontractible circle and a larger covering space.  In terms of
\begin{equation}
X^0 = (1 + \rho^2)^{1/2}  \cos \tau \,,\quad X^{D+1} = (1 + \rho^2)^{1/2}  \sin \tau \,,\quad
\rho^2 = \sum_{i=1}^D (X^i)^2 \,,
\end{equation}
the metric is
\begin{equation}
ds^2/L^2 = (1 + \rho^2) d\tau^2 + \frac{d\rho^2}{1 + \rho^2} + \rho^2 d\Omega^2_{S^{D-1}} \,.
\label{global}
\end{equation}
The range $0 < \tau < 2\pi$ gives the periodic space, and $-\infty < \tau < \infty$ gives global AdS space.  

Falling through the horizon would not seem to be relevant to the applications, since these always involve systems that are finite in size and duration: we turn the experiment off before anything reaches the horizon!  However, the global picture has several uses, some of which will be developed later: 1) it makes the $SO(D,2)$ conformal symmetry manifest; 2) it is dual to the CFT quantized on the space $S^{D-1}$ rather than $ {\mathbb R}^{D-1}$, if this is what interests us; 3) for this reason, it gives a mapping between the local operators of the CFT and the spectrum of states; 4) if we are interested in studying quantum gravity, the global spacetime is the natural place to formulate it.  Curiously, although the Poincar\'e patch is a small subset of the global space, the respective Wick rotations of the Poincar\'e $x^0$ and the global $\tau$ yield the same Euclidean version of anti-de Sitter spacetime.

Writing the $SO(D,2)$ as $\delta X^a = \epsilon^a\!_b X^b$, we can identify the action of the symmetries on the Poincar\'e coordinates: the $\epsilon^\mu\!_\nu$ are the Lorentz transformations, the  $\epsilon^\mu\!_U$ are the translations, the $\epsilon^U\!_U$ are the scale transformations, and the $\epsilon^\mu\!_V$ are the special conformal transformations.

\subsection{Scalar fields in AdS}

Consider a Klein-Gordon field, with Lorentzian action:
\begin{eqnarray}
S_0 &=& -\frac{\eta}{2L^{D-1}} \int dz\, d^dx\,( G^{MN} \partial_M \phi \partial_N \phi + m^2 \phi^2) 
\nonumber\\
&=&
-\frac{\eta}{2} \int \frac{dz\, d^dx}{z^{D+1}} \left( z^2 \partial_z \phi  \partial_z \phi
+ z^2 \eta^{\mu\nu} \partial_\mu \phi \partial_\nu \phi + m^2 L^2 \phi^2 \right) \,.
\end{eqnarray}
I've left in an arbitrary normalization constant $\eta$ for later use, but you can ignore it for now.
To discuss stability it is useful to define $\phi = z^{D/2}\psi$, $z = -\ln y$, so that
\begin{equation}
S_0 = -\frac{\eta}{2} \int{dy\, d^dx} \left(\partial_y \psi  \partial_y \psi
+ e^{-2y} \eta^{\mu\nu} \partial_\mu \psi \partial_\nu \psi + [m^2 L^2 + {\textstyle \frac14} D^2]  \psi^2 \right)  + \frac{\eta D}{4}  \int{d^dx}\, \psi^2 \Big|^{y = \infty}_{y = -\infty} \,.
\end{equation}
If we add a boundary term
\begin{equation}
S = S_0 + S_{\rm b} =  S_0 - \frac{ \eta\zeta}{2}  \int{d^Dx}\, \psi^2 |^{y = \infty} \label{bact}
\end{equation}
with $\zeta \geq \frac{D}{2}$,  and if $m^2$ satisfies the Breitenlohner-Freedman (BF) bound 
\begin{equation}
m^2 \geq -\frac{D^2}{4} \,,
\end{equation}
then the Hamiltonian is bounded below term-by-term and the system is stable~\cite{Breitenlohner:1982bm}.   In fact, the Hamiltonian can be organized into a sum of squares under the weaker condition $\zeta \geq \Delta_-$, defined in Eq.~(\ref{dpm}) below.

We see that a range of tachyonic masses is allowed.  If the mass-squared lies below the BF bound the free-field energy is unbounded below regardless of boundary terms, though it can be stabilized by higher order bulk terms.  As you will hear from some of the other speakers, this leads to interesting phase transitions.

The momentum space field equation is
\begin{equation}
z^{1+D} \partial_z (z^{1-D} \partial_z \phi_k) - (m^2 + k^2 z^2) \phi_k  = 0 \,.
\end{equation}
Near the $z=0$ boundary the $k^2$ term can be neglected and the solutions behave as
\begin{equation}
\phi_k \sim z^\Delta \,, \quad \Delta(\Delta - D) = m^2 L^2 \,.  \label{deltam2l2}
\end{equation}
There are two roots, 
\begin{equation}
\Delta_{\pm} = \frac{D}{2} \pm \left( \frac{D^2}{4} + m^2 L^2 \right) \equiv  \frac{D}{2} \pm \nu  \,,
\label{dpm}
\end{equation}
so the asymptotic behavior is
\begin{equation}
\phi(z,x) \sim \alpha(x) z^{\Delta_-} + \beta(x) z^{\Delta_+}  \,,\ z \to 0\,.
\end{equation}
We will consider the `standard' boundary condition $\alpha = 0$ and the `alternate' boundary condition~\cite{Klebanov:1999tb} $\beta =0$.  For the first of these the surface terms in the action and the equations of motion vanish due to the falloff of the fields.  For the second, the vanishing of the surface term in the equation of motion requires $\zeta = \Delta_-$; as we have noted, this gives a stable system in the tachyonic case.\footnote{One can think of the boundary condition in either of two ways: one can impose it on the fields from the start, or one can take free boundary conditions and let the equation of motion determine the boundary behavior.  These differ only by contact terms in the operator correlations. 
 \label{fixedfree}}  With the alternate boundary condition both the boundary term and the bulk action are divergent, but the divergences cancel (one should put the boundary at $z =\epsilon$ and then take $\epsilon \to 0$) provided that $\Delta_- > \frac{D}{2} - 1$; beyond this point the $k^2$ terms in the action diverge and only the standard boundary condition can be used.  We refer to the $z^{\Delta_+}$ solution as normalizable and the $z^{\Delta_-}$ solution as non-normalizable, though the later is normalizable in the more restrictive sense just discussed.
 
 \begin{enumerate}

\item[Ex.\ 5.] Extend the discussion to the degenerate case $\Delta_+ = \Delta_-$.  Show that the 
alternate quantization is consistent with conformal invariance in the generic case but not in the degenerate one.
\end{enumerate}

The general solution to the field equation is
\begin{equation}
\phi(z,x) = e^{i k \cdot x} z^{D/2} J_{\Delta - D/2}(qz) \,,
\end{equation}
where $q^2 = - k^2 < 0$ and $\Delta$ is whichever of $\Delta_\pm$ we are using.  These are plane-wave normalizable at the horizon $z\to\infty$, and correspond to a particle that comes out of the horizon, reflects off the boundary, and returns to the horizon.  If conformal symmetry is broken in such a way that the space is cut off near the horizon at some large value of $z$, the effective four-dimensional mass spectrum becomes discrete.

\subsection{Conformal field theories}

Now we discuss the field theory side of the duality.   Again, gauge/gravity duality is much more general than AdS/CFT, but the conformal case gives the cleanest examples to start from, so I will start with few general comments about CFT.  The basic observables are the correlation functions of local operators ${\cal O}(x)$.  Under scale transformations, an operator of dimension $\Delta$ transforms as
\begin{equation}
{\cal O}(x) \to \zeta^\Delta {\cal O}(\zeta x) \,.
\end{equation}
Thus scale invariance determines the form of the two-point function
\begin{equation}
\langle 0 | {\cal O}(x) {\cal O}(0) | 0 \rangle = \zeta^{2\Delta}\langle 0 | {\cal O}(\zeta x) {\cal O}(0) | 0 \rangle \
\Longrightarrow\ \langle 0 | {\cal O}(x) {\cal O}(0) | 0 \rangle  \propto \frac{1}{x^{2\Delta}} \,.
\end{equation}
Conformal invariance also determines the position-dependence of the three-point function, so that the basic data is a single coefficient $c_{ijk}$ for $\langle 0 | {\cal O}_i {\cal O}_j {\cal O}_k  | 0 \rangle$.  This is essentially the OPE coefficient
\begin{equation}
{\cal O}_i(x) {\cal O}_j(0) = \sum_k c_{ij}\!^k O(x^{\Delta_k - \Delta_i - \Delta_j}) {\cal O}_k(0) \,,
\end{equation}
the two-point function being used to raise and lower indices (also, the OPE of descendant fields is determined in term of their primaries).
The OPE has a nonzero radius of convergence, as it can be regarded under the state-operator mapping as simply the insertion of a complete set of states.  Then higher $n$-point functions can then be reduced to the two-point function using the OPE, and so are determined by the $c_{ijk}$.  The products in different channels have overlapping regions of convergence, so there is an associativity relation
\begin{equation}
\sum_k c_{ij}\!^k c_{kl}\!^m \sim \sum_k c_{il}\!^n c_{nj}\!^m \,,
\end{equation}
with appropriate functions of $x$ inserted.  There is always an energy-momentum tensor $T_{\mu\nu}$ which is traceless and has $\Delta = D$.  Its OPE coefficients are determined by the Ward identities.

This gives a purely algebraic description of a CFT, not referring directly to a Lagrangian.

\subsection{The $D=4$, ${\cal N} = 4$ theory}

The particular CFT of interest for the $AdS_5 \times S^5$ duality can be compactly defined as the dimensional reduction of the $D=10$ super-Yang-Mills theory, whose action is simply
\begin{equation}
S = \frac{1}{g_{\rm YM}^2} \int d^{10} x\, {\rm Tr}\left(- {\textstyle\frac{1}{2}} F_{MN}F^{MN} + i \bar\chi \gamma^M D_M \chi \right) \,. \label{n4act}
\end{equation}
To reduce to four dimensions, ignore the coordinates $x^M$ for  $M > 3$ and set the corresponding derivatives to zero.  The ten-dimensional gauge field separates into a four-dimensional gauge field and six scalars, 
\begin{equation}
A_M \to A_{\mu},\ \mu \leq 3\,,\ \ A_m,\ m \geq 4\,.
\end{equation}
The Majorana-Weyl spinor $\chi$, with 16 real components, separates into four $D=4$ Weyl spinors.  Further, the supersymmetry generators, also a Majorana-Weyl spinor in $D=10$, separates into four sets of $D=4$ generators. The scalars and spinors, like the gauge fields, are $SU(N)$ adjoints.  

The one-loop $\beta$-function vanishes by cancellation between the gauge field and matter contributions; we will give a nonperturbative agrument for this in Sec.~4.1.

It is sometimes useful to single out one supersymmetry and use ${\cal N}=1$ superfields.  There is one gauge multiplet and three chiral multiplets $\Phi_{1,2,3}$, and the superpotential is
\begin{equation}
W \propto {\rm Tr}(\Phi_1 [\Phi_2,\Phi_3]) \,.
\end{equation}
The scalar potential
\begin{equation}
V \propto \sum_{m,n}  {\rm Tr}|[A_m,A_n]|^2
\end{equation}
arises from the non-Abelian terms in the field strength in the dimensional ,reduction and as the sum of F-terms and D-terms in the ${\cal N}=1$ form.

Writing the prefactor of the action~(\ref{n4act}) as $N/\lambda$, the action is of order $N$ in the 't Hooft limit of large-$N$ fixed $\lambda$.  Each propagator then contributes a factor $N^{-1}$ and each interaction a factor $N$.  We are interested in expectation values of gauge-invariant operators,
\begin{equation}
{\cal O} = N\, {\rm Tr}(\ldots) \,.
\end{equation}
With this, an operator behaves like $N^{-1}$ times an interaction.  There is also a factor of $N$ for each index loop.  Filling the index loops in to make the faces of a simplex the total $N$-dependence is
\begin{equation}
N^{V - O +F-P} = N^{\chi-O}
\end{equation}
where $V,F,P,O$ count the vertices, faces, propagators, and operators, and $\chi$ is the Euler number of the simplex.  Thus for the ${\cal O}^4$ expectation value, the leading part of the fully disconnected term (the product of four one-point functions) has the topology of four two-spheres for $N^4$, the terms with two connected two-point functions has leading behavior $N^0$, and the fully connected amplitude has leading behavior $N^{-2}$For the fully connected contribution for $O$ operators the leading connected amplitude is $N^{2-O}$.

\sect{The dictionary}

In addition to classic references~\cite{GKP,W}, the review~\cite{D'Hoker:2002aw} is good for further details.  

\subsection{Symmetries}

The first check is a comparison of the symmetries of the two systems.  On the AdS side there are the $SO(4,2) \times SO(6)$ symmetries of the $AdS_5$ and $S^5$ spaces.  In the CFT, $SO(4,2)$ is the conformal group: we have already discussed at the end of Sec.~2.2 how the various symmetry transformations are realized.  The $SO(6)$ is the symmetry of the scalar field space $A_m$, with the four fermions transforming as the spinor representation.  Equivalently, the fermions are in the fundamental representation of $SU(4) = SO(6)$, and the six scalars in the antisymmetric product of two fundamentals.  On both sides of the duality this bosonic symmetry group is extended by supersymmetry to the superconformal $PSU(2, 2|4)$.  Again, there are gauge/gravity duals with much less symmetry, but it is useful to take account of it in this prototype example.

There is also an $SL(2,{\mathbb R})$ weak-strong duality on both sides, in the $D=4$, ${\cal N} = 4$ gauge theory and in the IIB string theory.

\subsection{Matching of states}

The duality implies a 1-1 mapping of the states on the two sides.  On the AdS side, since we are expanding in the loop parameter $L_{\rm P}/L$, we can count particle states at weak coupling.  On the CFT side, there is the usual isomorphism between states and operators: drawing a $(D-1)$-sphere around an operator, we can go to radial quantization and so we have an isomorphism between local operators and states on $S^{D-1}$.  Combining these, we can express the duality as a 1-1 mapping between the particle species in $AdS_{D+1}$ and the single-trace chiral primary operators in the CFT.  

The statement is that a field whose boundary behavior is $z^\Delta$ maps to an operator of dimension $\Delta$.  The mapping is simply that the scaled boundary limit of the bulk operator is the CFT operator,
\begin{equation}
{\cal O}(x) = C_{\cal O} \lim_{z\to 0} z^{-\Delta} \phi(x,z) \,. \label{opop}
\end{equation}
This fits with our original picture of the $z$ coordinate emerging from the size of the composite state: a local operator creates a state with zero size.  The constant $C_{\cal O}$ depends on convention, which we will set later.  The scale transformation in the bulk takes  $\phi(z,x) \to \phi(\zeta z, \zeta x) $ and so
\begin{equation}
{\cal O}(x) \to C_{\cal O} \lim_{z\to 0} z^{-\Delta} \phi(\zeta x,\zeta z) = C_{\cal O} \lim_{z\to 0} (z/\zeta)^{-\Delta} \phi(\zeta x, z) =  \zeta^{\Delta} {\cal O}(\zeta x) \,,
\end{equation}
which is the scale transformation of an operator of dimension $\Delta$.  Recall that the boundary behavior is related to the mass of the field by $m^2 L^2 = \Delta(\Delta-D)$, so that a given mass corresponds to two possible dimensions depending on the boundary condition taken.  The condition $\Delta > \frac{D}{2} - 1$ is precisely the lower limit on the dimension of an interacting scalar field in a unitary CFT.

We have focused on scalar fields, but the scalings work the same way for tensor fields if we use tangent indices rather than coordinate indices.
For example, every CFT contains the energy-momentum tensor, of dimension $D$, which maps to the graviton on the AdS side.  In terms of tangent space indices we then have for the metric perturbation $h_{\hat\mu \hat \nu} \sim az^D + b$, or $h_{\mu  \nu} = (z/L)^{-2} h_{\hat\mu \hat \nu} \sim az^{D-2} + bz^{-2}$.  Note that the larger behavior is the same as that of the AdS metric itself.

When the CFT has conserved currents, there is a corresponding gauge field in the bulk.  The dimension of a conserved current is $D-1$, so $A_{\hat\mu} \sim a z^{D-1} + b z$ and $A_\mu \sim a z^{D-2} + b$.  The constant term can be identified directly with a background gauge field in the \mbox{CFT}.  The $AdS_5 \times S^5$ theory has the $SO(6)$ global symmetry noted earlier, and correspondingly there are $SO(6)$ Kaluza-Klein gauge fields from the $S^5$.

There is an important distinction between states that would be massless in $D=10$ and get effective $D=5$ masses only due to the spacetime curvature, and those that are already massive in $D=10$.  The former have masses of order $1/L$ and so dimensions of order one.  In the $D=4$, ${\cal N} = 4$ case, there is enough supersymmetry in fact to guarantee that these dimensions are actually independent of the coupling.  This is because the maximum spin of such an operator is 2, which requires it to live in a BPS multiplet of the supersymmetry.  For duals with less supersymmetry the dimensions of such operators will be $O(1)$ at weak and strong coupling, but can depend on the coupling. 

Excited string states, on the other hand, have masses of order $\alpha'^{1/2}$ and so dimensions at strong coupling of order $\lambda^{1/4}$.  This is a striking prediction of the duality.  It is plausible that anomalous dimensions would become large at strong coupling, but there is no simple analytic argument for this particular behavior.  Results based on integrability, which are supposed to interpolate over all couplings, are consistent with it~\cite{Gromov:2009zb}, but there still seems to be some guesswork involved.  Eventually we might hope that the numerical approaches will reproduce it.

These large dimensions are essential to have a supergravity limit: in supergravity the maximum spin is two, so any operator of higher dimension must get a large anomalous dimension.\footnote{The higher spin theories constructed by Vasiliev may allow one to go beyond this in some cases.}  This requires very strong coupling, but it is really a stronger statement, as there are theories with couplings that are much larger than one but still without parametrically large anomalous dimensions for most operators.  Thus we can now elaborate on how we define strong coupling as a necessary condition for the duality: we need all stringy states, in particular those with spins greater than two, to get large anomalous dimensions. 

Focusing now on the specifics of the $D=4$, ${\cal N} = 4$, theory,
the gauge kinetic term $F_{\mu\nu}F^{\mu\nu}$ maps to the dilaton; it has dimension 4 at weak coupling, and (with appropriate additional pieces) it is BPS and so its dimension is exactly four. The topological term $F_{\mu\nu} \tilde F^{\mu\nu}$ maps to the RR scalar and has dimension 4.

The traces of products of scalars, ${\rm Tr} (A_m A_n \ldots A_p)$, give a large set of operators.  The linear combinations that are traceless on the $SO(6)$ indices map to shape fluctuations of the $S^5$, mixed with other modes.  The operator with $l$ scalars has exact dimension $l$.  The modes with traces, such as the Konishi operator ${\rm Tr} (A_m A_m)$ must map to excited string states. 

The superpotential perturbations~$\delta W(\Phi_{1,2,3})$ map to the perturbations of the harmonics of the NS and RR 3-form fluxes $F_{MNP}$, $H_{MNP}$ (which mix), and an operator with $m$ fields has dimension $m+1$.

There is a one-to-one correspondence between BPS operators of the CFT and supergravity modes in $AdS_5$, as required by the duality.  It might seem impossible for a four-dimensional field theory to have as many states as a ten-dimension supergravity theory, much less a string theory, and this would be many peoples' answer to Ex.~1, e.g.~\cite{Penrose}.  The key is that there are a lot of states due to the large matrices.  We have seen that the Kaluza-Klein excitations come from traces of many operators, and these traces are all independent as $N \to \infty$.  At one level it seems trivial that one can encode anything in a large-$N$ matrix; what is remarkable is that the codebook is just the ${\cal N} = 4$ path integral.

This also provides another nice illustration of the duality working nonperturbatively in $N$.  At finite $N$ the the traces with more than $N$ operators are not independent, and so there must be a cutoff on the momentum parallel to the $S^5$.  One can make harmonics up to $l = N$, corresponding to momenta of order $N/L \sim N^{3/4}/L_{\rm P}$.  It is might seem wrong that the cutoff is much larger than the Planck scale, but there is no bar to boosting a particle up to highly super-Planckian momenta.  The cutoff arises rather from a curious phenomenon~\cite{McGreevy:2000cw}: the interaction of the particle with the flux on the $S^5$  causes it to blow up to the point that it can no longer fit in the space.

Rather than counting supergravity states, we could consider the total partition function of the theory.  In the gravity picture this is dominated by the black 3-brane, whose entropy per unit volume is
\begin{equation}
s_{\rm black\,D3} = \frac{\pi^2}{2} N^2 T^3 \,,
\end{equation}
to be compared with the free-field entropy of the CFT,
\begin{equation}
s_{\rm free\,CFT} = \frac{2\pi^2}{3} N^2 T^3 \,.
\end{equation}
These agree up to the famous factor of 3/4~\cite{Gubser:1996de}: the duality is saying that as the coupling is increased from zero to infinity in the gauge theory, there is this small shift in the density of states.
The agreement of the temperature dependences is confirming that the AdS theory is behaving like a four-dimensional CFT, and the agreement of the $N$-scaling is another satisfying check.  However, there could have been a nontrivial dependence on the 't Hooft coupling $\lambda$, so that the weak and strong coupling entropies could have differed for example by a power of $\lambda$.  There is no independent check of this result, so for now it is a prediction of the duality rather than a check. 

\subsection{Correlators I}

I am going to take the operator relation~(\ref{opop}) as the starting point for the dictionary.  First, I would like to dispel a myth, that the AdS/CFT dictionary is somehow naturally Euclidean, and that there are difficulties with extending it to Lorentzian spaces.\footnote{If anything the reverse is true, since there is still no comprehensive treatment of the contour for the conformal factor in the path integral for Euclidean gravity.}  In the first place, when we compare the spectra of the Hamiltonian on the two sides, this makes no reference to any signature, since the difference is just whether we evolve forward with $e^{-iHt}$ or $e^{-H\tau}$.  Secondly, the basic relation~(\ref{opop}) holds equally in any signature.  It is true that in Lorentzian signature there are several correlators of interest (time-ordered, advanced, retarded), and the prescriptions for their calculation, at finite temperature, are intricate.  This will be discussed in the lectures by Son; for a recent overview see Ref.~\cite{Skenderis:2008dg}.

But these questions are not issues for the duality, which via~(\ref{opop}) just relates any given correlator in the CFT with the same correlator in the AdS space.

I want to calculate the time-ordered 2-point correlator in several ways.  First, the propagator of $\phi$ in the bulk of $AdS_{D+1}$ is
\begin{equation}
\langle 0 | {\rm T}\, \phi(z,x) \phi(z',x') | 0 \rangle = \frac{1}{\eta} G_\Delta(\xi) \,,
\end{equation}
where $\Delta = \Delta_+$ or $\Delta_-$ according to the choice of boundary condition.
Here 
\begin{equation}
\xi = \frac{2 z z'}{z^2 + z'^2 + (x-x')^2}  
\end{equation}
is the unique conformal invariant that can be constructed from the two positions.  It is a measure of the distance between them, approaching 1 when they become coincident and 0 when they are far apart.  Emphasizing again the signature of spacetime, this is equally valid in the Euclidean and Lorentzian theories, with
\begin{equation}
(x-x')^2 \equiv_{\rm E} \sum_{i = 1}^D (x^i - x'^i)^2 \equiv_{\rm M} (i+\epsilon)^2 (x^0- x'^0)^2 
+ \sum_{i = 1}^{D-1} (x^i - x'^i)^2 \,,
\end{equation}
and we are working in the Poincar\'e patch.

The propagator $G_\Delta(\xi)$ is hypergeometric, and Ref.~\cite{D'Hoker:2002aw} is a good source for more details.  When the points are far apart, and this includes the case that $z$ and/or $z'$ approach the boundary with $x - x'$ fixed and nonzero, then
\begin{equation}
G_\Delta(\xi) \to \frac{C_{\Delta}}{2\Delta-D} (\xi/2)^\Delta  \,,\quad C_{\Delta} = \frac{\Gamma(\Delta)}{\pi^{D/2} \Gamma(\Delta-D/2)} \,.
\end{equation}
In particular, it behaves as $z^\Delta z'^\Delta$ as required by the boundary conditions.  Thus we can form the limit~(\ref{opop}) and conclude that
\begin{equation}
\langle 0 | {\rm T}\, {\cal O}(x) {\cal O}(x') | 0 \rangle =  \frac{C_{\Delta}C^2_{\cal O}}{\eta(2\Delta-D)} (x - x')^{-2\Delta} \,.
\end{equation}
Of course the 2-point correlator is determined up to normalization by scale invariance, but we are just warming up for the later lecturers, who will look at the 2-point function after turning on temperature, densities, or background fields, where it captures much interesting physics.

In some cases the normalization of the 2-point function is interesting, but here it is just the product of a bunch of conventional factors.  We now introduce a standard convention.  If we take just the primed field to the boundary we get
\begin{equation}
\langle 0 | {\rm T}\, \phi(z,x) {\cal O}(x') | 0 \rangle =  \frac{C_{\Delta}C_{\cal O}}{\eta(2\Delta-D)} {\left( \frac{z}{z^2 + (x-x')^2 }\right)}^{\Delta}\,.
\end{equation}
If we now take $z \to 0$, the correlator goes to zero pointwise except at $x = x'$, and its value integrated $d^D x$ is $z^{D-\Delta} C_{\cal O} / \eta(2\Delta-D)$.  It is therefore conventional to set $C_{\cal O} = \eta(2\Delta-D)[N^{-1}]$ (ignore for now factors in square braces!), so that
\begin{equation}
\langle 0 | {\rm T}\, \phi(z,x) {\cal O}(x') | 0 \rangle \stackrel{z \to 0 }{ \to} z^{D-\Delta} \delta^{D}(x-x') [N^{-1}] \,. \label{nonorm}
\end{equation}
At this point I am going to switch to a Euclidean metric, not because I have to but because it is conventional in this subject (also, the Lorentzian form would have a factor of $-i$ in Eq.~(\ref{nonorm}) and compensating factors elsewhere); anyway, I hope I have already made my point about the duality being just fine in Lorentizian signature.  Note that this also requires that I flip the sign of the action due to standard conventions.
Noting that $D - \Delta_+ = \Delta_-$ (and $D - \Delta_- = \Delta_+)$, the operator ${\cal O}$ can be regarded as a delta-function source for the mode that was previously set to zero.  

With this normalization we also have
\begin{eqnarray}
\langle 0 | {\rm T}\, {\cal O}(x) {\cal O}(x') | 0 \rangle &=&  \frac{\eta(2\Delta - D)C_{\Delta}}{(x - x')^{2\Delta}} [N^{-2}]\,, \nonumber\\
\langle 0 | {\rm T}\, \phi(z,x) {\cal O}(x') | 0 \rangle &=& C_{\Delta} {\left( \frac{z}{z^2 + (x-x')^2 }\right)}^{\Delta} [N^{-1}]\,. \label{normed}
\end{eqnarray}
The constant $\eta$ arises from the normalization of the Klein-Gordon action.\footnote{Noting the discussion of operator normalization in Sec.~2.5, we see that $\eta$ must be of order $N^0$.  On the other hand, it is often convenient to normalize all closed string fields like the graviton, with a $1/G$ in the action, so $\eta \sim L^{D-1}/G_{D+1} \sim L^8 / G_{10} \sim N^2$.  In this case one must include the factors in braces.  In the remainder of these notes we use only the $\eta \sim N^0$ convention.}

The relation~(\ref{nonorm}) allows us to connect with the more standard way to write the dictionary.  Noting that $D - \Delta_+ = \Delta_-$ (and $D - \Delta_- = \Delta_+)$, the operator ${\cal O}$ can be regarded as a delta-function source for the mode that was previously set to zero.  In particular, if we introduce into the CFT path integral a factor
\begin{equation}
e^{\int d^dx\, j(x) {\cal O}(x)} \,,
\end{equation}
then we now have
\begin{equation}
\alpha(x) \to j(x)\ \ \mbox{(standard quantization)}\,;\quad \beta(x) \to j(x)\ \ \mbox{(alternate quantization)}\,. \label{jbound}
\end{equation}
In other words, the coefficient of the fixed ($z^{D-\Delta}$) mode is that with which the operator $\cal O$ is added to the CFT, while the coefficient~(\ref{opop}) of the $z^\Delta$ quantized mode is the expectation value of $C_{\cal O}^{-1}\cal O$.

Evaluating the bulk path integral then gives a generating functional
\begin{equation}
\langle 0 | {\rm T}\, e^{\int d^D x\, j(x) {\cal O}(x)}|  0 \rangle = Z_j \to e^{-S_{\rm cl}}\,.
\end{equation}
Here $Z_j$ is the bulk path integral with boundary condition~(\ref{jbound}), and in the last form we have evaluated it in the semiclassical approximation, extremizing with respect to the field with given boundary conditions.  Using Eqs.~(\ref{normed}, \ref{jbound}), the extremum is
\begin{equation}
\phi_{\rm cl}(z,x) = C_{\Delta} \int d^Dx' \, j(x') \left( \frac{z}{z^2 + (x-x')^2 }\right)^{\Delta}
\end{equation}
 This corresponds to the planar approximation in this CFT, but is also exact for the quadratic theory being studied.  This is the form in which the dictionary is often given.

Integrating by parts, the classical bulk action can be written as a surface term (again, all signs flipped from the previous discussion)
\begin{equation}
S_{0,\rm cl} = -\frac{\eta}{2L^{D-1}} \int dz\, d^Dx\,\phi_{\rm cl} ( \Box - m^2) \phi_{\rm cl} 
+ \frac{\eta}{2} \epsilon^{1-D} \int d^dx \,\phi_{\rm cl}(\epsilon, x) \partial_\epsilon \phi_{\rm cl}(\epsilon, x) \,,
\end{equation}
the first term vanishing by the equation of motion.  This is not the whole story because of the need for a boundary term in the action.
To regulate potential divergences, we temporarily move the boundary in to a small value $z = \epsilon$.  The boundary term must be such that the boundary terms in the variation of the action vanish for variations that respect the boundary conditions; this is in order to have a good variational principle.\footnote{This may still leave some freedom in the choice of the boundary action, which would just correspond to changes of convention such as operator redefinition.  Also, finiteness  of the total bulk plus boundary action as $\epsilon \to 0$ might be desirable, but the divergences are local contact terms and so easily subtracted by hand.}  One good choice of boundary action for the standard quantization is
\begin{equation}
S_{\rm b} =  \frac{ \eta\Delta_-}{2} \epsilon^{-D}  \int{d^Dx}\, \phi^2(\epsilon, x) \,,
\end{equation}
which also guarantees stability as discussed earlier.
Then 
\begin{equation}
\delta(S_0 + S_{\rm b}) = -{\eta} \epsilon^{-D} \int d^Dx \, \delta\phi(\epsilon, x) (\epsilon\partial_\epsilon  - \Delta_-)\phi(\epsilon, x) \,.
\end{equation}
The boundary condition fixes $\alpha(x)$, so $\delta\phi(\epsilon,x)$ has only a $\delta\beta$ term, and then only the cross term with $\alpha$ in the second field survives at the boundary, and this is annihilated by  $\epsilon\partial_z  - \Delta_-$.  One can also check that the potentially divergent $\alpha^2$ terms cancel for the same choice of boundary term.  In all
\begin{equation}
S_{\rm cl}  =  -\frac{\eta}{2} \epsilon^{-D} \int d^Dx \,\phi_{\rm cl}(\epsilon, x) (\epsilon \partial_\epsilon - \Delta_-)\phi_{\rm cl}(\epsilon, x) \,.
\end{equation}
Inserting $\phi_{\rm cl}$ leaves a convolution to do~\cite{W,mcgnotes}, but it is not hard to deduce the answer.  The $\epsilon \partial_\epsilon - \Delta_-$ kills the $\alpha$ part of the second $\phi_{\rm cl}$, so we must get precisely the $\alpha$ part of the first $\phi_{\rm cl}$ in order that the term survive at the boundary, and this is just $j(x)$.  So
\begin{eqnarray}
S_{\rm cl}  &=&  -\frac{\eta}{2} \epsilon^{-\Delta} \int d^Dx \,j(x) (\epsilon \partial_\epsilon - \Delta_-)\phi_{\rm cl}(\epsilon, x)   \nonumber\\
&\to&  -\frac{\eta}{2} (2\Delta - D) C_{\Delta} \int d^Dx  \int d^Dx' \,j(x) j(x') \frac{1}{ (x-x')^{2\Delta }}\,,
\end{eqnarray}
correctly generating the earlier result.  The extensions to the alternate quantization, and to the more general nonconformal boundary condition $\alpha = f\beta + j$, are left as exercises.  The solution for the latter is given below Eq.~3.22 of Ref.~\cite{Heemskerk:2010hk}.

\subsection{Correlators II}

For gauge fields and metric perturbations it is much the same, except that the symmetries give a natural normalization.  We have noted in Sec.~3.2 that the non-normalizable mode of a gauge field scales as $z^0$, so we can interpret its limit directly as a gauge field in the CFT,
\begin{equation}
\lim_{z\to 0} A_\mu(z,x) = A_{\rm b\mu}(x) \,.
\end{equation}
This identification is necessary in order that gauge transformations act consistently on charged operators.
This mode couples to the corresponding conserved current in the CFT as $\int d^Dx\, A_{\rm b\mu}(x) j^\mu(x)$.  Similarly the nonnormalizable mode of the metric has the same scaling as the AdS metric, and so we interpret
\begin{equation}
\lim_{z\to 0} z^2 h_{\mu\nu}(z,x) = h_{\rm b\mu\nu}(x) \,, 
\end{equation}
which couples to the energy-momentum tensor in the CFT as $\frac{1}{2} \int d^Dx\, h_{\rm b\mu\nu}(x) T^{\mu\nu}(x)$.  

Since later lecturers will be looking the the $TT$ correlator, let us look at one component of this, at zero temperature.  We excite $h \equiv h_{12}$ with momentum $k$ in the 3-direction.  So far we have worked in position space, but with less symmetry there will not be a simple coordinate form and so we go to momentum space.  The relevant terms in the Euclidean action in $AdS_5$ are
\begin{equation}
\frac{L^3}{4 \hat L_{{\rm P},5}^{3}} \int \frac{ d^4x\,dz}{z^{d-1}} (\partial_z h \partial_z h + k^2 h^2) \,.
\end{equation}
This is exactly like the massless Klein-Gordon field.  The usual gravitational extrinsic curvature term has been included implicitly to make the action first order~\cite{Gibbons:1976ue}, but no additional surface term is needed because $\Delta_- = 0$.

The extremum can again be written as a surface term,
\begin{equation}
S_{\rm cl} =  -\frac{L^3}{8 \hat L_{\rm P}^{3}} \epsilon^{-3} \int d^dx\,  \partial_ \epsilon (h^2(\epsilon) ) \,.
\end{equation}
The classical solution that approaches $h_{\rm b}$ at the boundary and remains bounded at $z= \infty$ is 
\begin{equation}
h_{\rm cl}(z) =  h_{\rm b} \frac{k^2 z^2}{2} K_{2}(kz) \,.
\end{equation}
Expanding the Bessel function at small $z$,
\begin{equation}
h_{\rm cl}(z) = h_{\rm b} \left\{ 1 + O(k^2 z^2) -  \frac{k^4 z^4}{16}  \ln k +  O(k^4 z^4 \ln z) + \ldots \right\}
\end{equation}
There is a quadratically divergent $\epsilon^{-2} k^2$ piece; this is analytic in $k^2$ and so is a contact term.  There are also $\ln \epsilon$ and finite $k^4$ contact terms.  The term of interest ir $k^4 \ln k$, leading to 
\begin{equation}
S_{\rm cl} =  \frac{h^2_{\rm b} L^3 k^4 \ln k}{16 L_{\rm P}^{3}} \,,
\end{equation}
and
\begin{equation}
\langle 0 | T_{12}(k) T_{12}(0) | 0 \rangle =  \frac{L^3 }{8 \hat L_{\rm P}^{3}} k^4 \ln k \,.
\end{equation}
Using $\hat L_{\rm P,5}^{3} = \hat L_{\rm P,10}^{8}/V_{S^5}$ where $V_{ S^5} = \pi^3 L^5$, this becomes~\cite{GKP}
\begin{equation}
\langle 0 | T_{12}(k) T_{12}(0) | 0 \rangle =  -\frac{\pi^3 L^8 }{8 \hat L_{\rm P,10}^{8}} k^4 \ln k
= - \frac{N^2}{32 \pi^2} k^4 \ln k \,.
\end{equation}
In position space this goes as $1/x^8$ as it must for an operator of dimension $4$.  The central charge $c$ is defined as $-8\pi^2$ times the coefficient of $k^4 \ln k$, i.e. $N^2/4$.  It must be independent of the coupling because it is related by supersymmetry to the $SO(6)$ R-anomaly, and indeed agrees with the weak coupling result \cite{Gubser:1997se}.

\sect{Breaking symmetries}

In this final section I want to generalize the example we have been studying in various ways.  Two challenging goals are to construct duals to landscape AdS states, and to construct non-Fermi liquids.  We will not reach these, but I will discuss some issues.  The landscape states, as mentioned before, have the property that the compact dimensions can be much smaller than the AdS space, and also that one can uplift them to higher AdS states, and also dS, by breaking supersymmetry (though as Tom Banks emphasizes, the uplifted states may not be visible in the AdS construction).  The non-Fermi liquids represent a set of observed strongly coupled fixed points that have defied a clear description.

In the applications to condensed matter physics there are two approaches being taken.  One is to postulate a gravitational theory with some set of fields, implying that the CFT has the corresponding operators.  There will always be the gravitational field, corresponding to $T_{\mu\nu}$, and there may be some gauge fields in the bulk, mapping to symmetry currents in the CFT, and also some charged matter fields, mapping to charged operators of various dimensions.  In this case there is no specific Lagrangian in the CFT, the results apply to any theory with the given set of operators: once one assumes some specific operators (and their dimensions and maybe some couplings) the dual allows one to calculate how the system responds to temperature, electric and magnetic fields, and other perturbations.  This has been instructive, and led to interesting work on non-Fermi liquids which Hong Liu will discuss, but it would seem more satisfying to start from some brane system and deduce the Lagrangian and its gravity dual.  This top-down approach has been challenging, as I will describe.

\subsection{The Coulomb branch}

Starting from the $D=4,\ {\cal N}=4$ example, there is a way to break the conformal symmetry without changing the theory, but just going to a different vacuum.  The scalar potential $\sum_{m,n}  {\rm Tr}|[A_m,A_n]|^2$ vanishes whenever the six matrices $A_m$ commute, but they can otherwise take any values.  This corresponds to separating some or all of the 
D3-branes in the six transverse directions of the Poincar\'e picture.  There is no force between them due to supersymmetry; the potential gets no quantum corrections.  Separating the D3-branes breaks the gauge invariance to a block for each set of coincident D3-branes, a product of $U(1)$ and $SU$ factors, and it breaks the scale and conformal invariance because some of the branes are sitting at nonzero $r$.  A fairly recent discussion of this subject is Ref.~\cite{Skenderis:2006di}.
 
From the point of view of the applications, this is more of a nuisance perhaps: most of them don't have fundamental scalars but in the duality these come along due to supersymmetry.  And, when we break the supersymmetry, it may be that the vacuum that we want (where the scalar vacuum expectation values are zero) is no longer stable.  The top-down construction of a Fermi sea has been a goal of mine for a long time because of the possibility of constructing a non-Fermi liquid, but these constructions often seem to have this problem.

We can use these vacua to give a modern nonperturbative proof of the conformal invariance of the ${\cal N}=4$ theory.  If we make all of the vevs distinct, the gauge group is $U(1)^{N-1}$ and the low energy effective fields are just the gauge fields and their scalar $A_m$ and fermionic partners.  The low energy effective action for the gauge fields is
\begin{equation}
-\sum_i \frac{1}{4g_i^2} F^{\vphantom \mu}_{i\mu\nu}F_i^{\mu\nu}  \,. \label{effab}
\end{equation}
If the couplings run, then the $g_i$ should be evaluated at a scale of order the masses of the higged gauge bosons, which are proportional to the $A_m$.  But the action will then depend on the scalars, and this is forbidden by the ${\cal N}=4$ supersymmetry.

\subsection{Renormalization group flows}

If we start with the $D=4,\ {\cal N}=4$ theory and perturb the Hamiltonian by a relevant operator, $\Delta < D$, the effect of the perturbation becomes large at low energies.  This is readily studied in the dual theory: adding $g \int d^Dx\, {\cal O}(x)$ to the action for some operator $\cal O$ shifts the corresponding fixed mode.  For an operator of dimension $\Delta$ this mode behaves as $z^{D-\Delta}$ which is just right: for relevant operators, $\Delta < D$, their effect grows as we move away from the boundary.\footnote{If we try to add an irrelevant operators $\Delta > D$, the effect grows as we approach the boundary, the nonlinearities become large, and generically we will lose the AdS geometry and the duality.}  

The interesting question is, what happens at low energies where the perturbation gets large?  The two broad classes of possibility are that we might flow to a new conformal theory, or that everything just gets massive.  We can find examples of each within the ${\cal N}=4$ theory.  A perturbation that preserves ${\cal N}=1$ supersymmetry is to add a mass term for one or more of the chiral superfields,
\begin{equation}
\delta W = \sum_{\alpha = 1}^3 m_\alpha \Phi^2_\alpha \,.
\end{equation}
This is a nice example because the low energy physics depends in an interesting way on how many of the fermions are massive.

First on the field theory side, for one nonzero mass there are field theory arguments that the system flows to a new conformal fixed point~\cite{Leigh:1995ep}.  For two equal nonzero masses, there is a ${\cal N} = 2$ supersymmetry and the massless sector is just the pure ${\cal N} = 2$ super- Yang Mills, whose low energy physics is the $SU(N)$ version of Seiberg-Witten~\cite{Douglas:1995nw}, where the massless fields are described by a $U(1)^{N-1}$ effective theory.  (For two unequal masses I am not sure what happens, but I'm guessing that it flows to the same fixed point).  For three nonzero masses~\cite{Polchinski:2000uf}, there are actually many supersymmetric vacua: the superpotential equations 
\begin{equation}
\label{su2rep}
m\Phi_\alpha
= \epsilon_{\alpha\beta\gamma} [\Phi_\beta, \Phi_\gamma]
\end{equation}
give a vacuum for each $N$-dimensional representation of $SU(2)$.  The vacuum corresponding to the trivial solution $\Phi_\alpha = 0$ is expected to be confining, while all the others have the gauge group broken to some subgroup containing non-Abelian factors that confine and always some unbroken $U(1)$'s.

On the gravity side, the mass perturbation corresponds to a perturbation of the 3-form flux on the $S^5$.  One can analyze this in terms of an effective Lagrangian for just this mode coupled to gravity.  This seems like a cheat, because this is not a real effective field theory, there is no symmetry that allows us to truncate to just this mode, but there is some supergravity magic that allows one to extend any solution of this truncated system to the full theory.  The effective potential looks like the curve $x^3 - x$, with one maximum and one minimum.  The $AdS_5 \times S^5$ theory actually sits at the maximum, which looks unnerving but is fine because of the BF bound: the field can't roll down in time, but it can roll in the radial direction, ending up at large $z$ at the minimum, which is a new CFT.  The overall geometry in the infrared is $AdS_5 \times X$ where $X$ is some deformed $S^5$ with three-form flux~\cite{Khavaev:1998fb}.

One can repeat this with two masses, but now the field flows off in a different direction, to infinity.  The supergravity magic allows this to be lifted to a ten-dimensional solution which is now singular, but the singularities have a nice interpretation: they are just $N$ D3-branes~\cite{Pilch:2000ue,Buchel:2000cn}.  The $z$-coordinate (defined so that the metric is $L^2 \eta_{\mu\nu}/z^2 +\ldots$) is cut off at a finite upper limit, meaning that there is a lower bound on the energies $P_0 = L {\cal E}/z$.  However, there are massless states on the D3-branes, giving just the expected $U(1)^{N-1}$ theory.

When masses are turned on for all three superfields, one can try to repeat the above effective Lagrangian strategy.  It again shows the geometry ending before $z = \infty$, but now the five-dimensional solution~\cite{Girardello:1999bd} lifts to a ten-dimensional solution with an unphysical singularity, it is not quite right.  One must work in fully ten-dimensional terms.  With some guesswork~\cite{Polchinski:2000uf}, guided by the field theory, one can find a sensible solutions where the geometry ends at some finite $z$, and there are explicit NS5- and/or D5-branes, with D3-branes dissolved in them.  The D5's correspond to non-commuting coordinate matrices $A_m$ for the original D3's, as suggested by~(\ref{su2rep}): this is the Myers effect~\cite{Myers:1999ps}, where in a background field it can become energetically favorable for the D3 coordinates not to commute.  For the NS5's, there is some strong-coupling dual of this effect, which has no such classical description.  The confining vacuum is the one with just a single NS5-brane.

To talk about confinement we need another observable, which measures the force between quark-like sources~\cite{Rey:1998ik}.  This is simply the energy of a string that starts and ends on the the sources (one can give a longer justification, but this is just part of 't Hooft's large-$N$ story, where the string corresponds to a tube of color flux).  Now, the effective tension of such a string, as seen in the gauge theory, is $\mu(z) = 2\pi L^2/\alpha' z^2$, since the tension seen by a local inertial observer is $2\pi/\alpha'$.  In the conformal theory, as one move the sources apart, the string can drop down closer and closer to the horizon and reduce its tension indefinitely.  One then gets $V(r) \propto 1/r$ (as one must in a conformal theory, though it is interesting that at strong coupling the potential goes as $\lambda^{1/2}$ rather than as $\lambda$).

In the confining vacua, there is a maximum $z$ and so a minimum tension $2\pi L^2/\alpha' z_{\rm max}^2$.  There are many other realizations of confinement in AdS/CFT~\cite{W,Klebanov:2000hb, Maldacena:2000yy}; others are simpler on the gravity side, but generally more complicated on the CFT side.  In the Coulombic vacua, as with two masses, strings hang down from each source and end on a D3-brane, where they source an explicit flux that can spread out coulombically.

\subsection{Multitrace RG flows}

If the CFT has a scalar operator of dimension $\frac{D}{2} - 1 < \Delta < \frac{D}{2}$, this is described in the bulk by the alternate quantization in which the larger mode $\alpha$ fluctuates and the smaller mode $\beta$ is fixed.  There are actually no such operators in the ${\cal N}=4$ theory, but there are in many other examples.

In this case the double-trace operator ${\cal O}^2$ is relevant~\cite{Witten:2001ua,Berkooz:2002ug}: 
\begin{equation}
\Delta_{{\cal O}^2} = 2\Delta + O(1/N^2) \,,
\end{equation}
because the large-$N$ factorization means that any graph that connects the two single-trace operators is down by $1/N^2$.  Thus if we add this to the action there will be some RG flow.
To see where it goes, write~\cite{Berkooz:2002ug}
\begin{eqnarray}
Z_j(g) &=& \left\langle e^{\int d^Dx\,\{ -g {\cal O}^2(x)+ j(x) {\cal O}(x) \}} \right\rangle_{\beta=0} \nonumber\\
&=& \int {\cal D}\sigma \, \left\langle e^{\int d^Dx\,\{ - \sigma^2(x)/4g + (j(x) + i\sigma(x)) {\cal O}(x) \}}
\right\rangle_{\beta=0} \nonumber\\
&=& \int {\cal D}\sigma \, \left\langle e^{- \int d^Dx\, \sigma^2(x)/4g }
\right\rangle_{\beta= j + i\sigma}\nonumber\\
&=& \int {\cal D}\sigma \,  e^{- \int d^Dx\, \sigma^2(x)/4g }
 \left\langle1\right\rangle_{\beta= j +i \sigma}
 \label{aux}
\end{eqnarray}
In the first line we have written the generating functional for expectation values of $\cal O$ with the perturbation $g{\cal O}^2(x)$ in the Hamiltonian,\footnote{It was asked whether $g$ must be positive.  Naively, negative $g$ leads to instability, but recently it has been shown~\cite{Faulkner:2010gj} that nonlinear effects of backreaction can stabilize the system.}
 and in the second line we have introduced an auxiliary field to relate this to the theory with just the single-trace perturbation.  Now let use integrate out the auxiliary field.  In the planar approximation we can just find the saddle point,\footnote{Loop corrections for auxiliary fields are generally uninteresting, being dominated by the cutoff scale for dimensional reasons, and so absorbable into redefinitions.}
\begin{eqnarray}
\sigma(x) &=& -2g \partial_\sigma Z_{\beta= j + i\sigma} \nonumber\\
&=& 2gi  \left\langle {\cal O}\right\rangle_{\beta= j + i\sigma} \nonumber\\
&=& 2gi C_{\cal O} \alpha_{\beta=j +i\sigma} \,.
\end{eqnarray}
In the second line we have used the fact that differentiating with respect to $j$ is the same as with respect to $\sigma$, and pulls down a $\cal O$.  In the last line we have used the basic dictionary~(\ref{opop}).  Eliminating $\sigma$
\begin{equation}
\beta(x) = j(x) - 2g C_{\cal O} \alpha(x) \,.
\end{equation}
In other words, the earlier pure boundary conditions are now replaced by a mixed condition with a linear relation between $\alpha$ and $\beta$~\cite{Witten:2001ua,Berkooz:2002ug}.  In terms of the field we have
\begin{equation}
\phi(z,x) = z^{D-\Delta} \{ j(x) - 2g C_{\cal O} \alpha(x) \} + z^{\Delta} \alpha(x)   \,.
\end{equation}
At high energy $z \to 0$ the $z^{D-\Delta}$ term vanishes (we are in the regime $\Delta <  D/2$) and we have the alternate quantization.  At low energy the $z^{D-\Delta}$ term dominates and we have the usual quantization.  Thus we flow from one to the other.

By the way, in the middle two lines of Eq.~(\ref{aux}) we see an interesting construction: a field $\sigma$ coupled to the boundary of the supergravity.  In this case the field is auxiliary, but we could consider more general actions for $\sigma$.  This corresponds to coupling some strongly coupled field theory, with a gravity dual, to another field theory (generally weakly coupled) with an explicit Lagrangian description.

\subsection{Orbifolds}

Tom Banks discussed some features of these, let me add a little.  If we identify the ${\mathbb R}^6$ space transverse to the D3-branes under some discrete group $\Gamma \subset SO(6)$ then the origin becomes a singularity.  If we then take the near horizon limit, then we get a new duality~\cite{Kachru:1998ys}. On the supergravity side $S^5$ is replaced by $S^5/\Gamma$.  On the gauge side, the orbifolding retains only the $\Gamma$-invariant fields from the original gauge theory, leaving a quiver gauge theory~\cite{Douglas:1996sw}, e.g.~for $\Gamma={\mathbb Z}_k$ we get $SU(N)^k$, with bifundamental matter that depends on how $\Gamma$ acts on the different planes in ${\mathbb R}^6$.  The surviving supersymmetry can be ${\cal N} = 2, 1,$ or 0.

In the nonsupersymmetric cases, Tom mentioned that at weak coupling conformal invariance is destroyed by the flow of double-trace interactions.  The pathology takes a different form at strong coupling.  If a nonsupersymmetric element of $\Gamma$ has a fixed point there will be a localized tachyon, with the result that a hole develops and consumes the space~\cite{Adams:2001sv}.  If there is no fixed point the same process occurs via tunneling~\cite{Horowitz:2007pr}.  These are examples of the general phenomenon that nonsupersymmetric states in string theory are almost always unstable on long enough time scales.

There is an interesting generalization where $SU(N)^k$ is replaced by $SU(N_1) \times \ldots \times SU(N_k)$.  On the supergravity side this corresponds to wrapping higher dimensional branes (D5 in particular) on cycles hidden in the singularity.  The different gauge factors now have different nonzero $\beta$-functions and the couplings flow.  Figuring out where the flow ends up has been a rich subject~\cite{Klebanov:2000hb}.

\subsection{Non-spherical horizons. II}

If we combine orbifolding and RG flow we get something new.  Some supersymmetric orbifolds have relevant operators in the twisted sector, corresponding to smoothing the fixed point.  By turning these on one gets to new manifolds, such at the famous conifold~\cite{Klebanov:1998hh}.  Here the $S^5$ is replaced by a manifold $T^{1,1}$.  In fact one gets a solution to the supergravity field equations with just metric and five-form if $S^5$ is replaced by any Einstein space ($R_{mn} = c g_{mn}$ with $c$ a positive constant).  Supersymmetry requires a more restrictive Einstein-Sasaki condition~\cite{Kehagias:1998gn}.  The dual gauge theory can be determined in many cases by a combination of orbifolding, RG flow, moving to the Coulomb branch, and so on.  The toric case, where $X$ has a $U(1)^3$ symmetry, is understood best ~\cite{Franco:2005sm}.  There has been recent progress on the more general case~\cite{Aspinwall:2010mw}.

\subsection{Nonconformal branes~\cite{Itzhaki:1998dd}}

Suppose we try to repeat the near-horizon limit for D2-branes.  Here the gauge theory is superrenormalizable: at high energy it is weak, and the dimensionless coupling $g_{\rm YM}^2 N /{ E}$ becomes small.  In the IIA supergravity theory, the curvature is small in some range, but becomes large in string units near the boundary.  Thus we have the same complementarity as for D3-branes, but at different scales within a given theory rather than by varying the coupling.  The superrenormalizable gauge theory is the Lagrangian description, and the gravitational theory describes the low energy physics.  

There is more to the story, though.  The growth of the dimensionless gauge coupling is also reflected in the blowup of the dilaton at large $z$.  In this regime one must lift to M theory, and the extreme low energy limit is the M2-brane geometry.  Thus the extreme IR of $D=3,\ {\cal N}=8$ gauge theory provides one Lagrangian description of this geometry.  I like this $D=3$ theory as an `existence proof' of AdS/CFT as a theory of quantum gravity, because we can regulate it on the lattice without exact supersymmetry, and the superrenormalizability makes it easy to see the restoration of the symmetry in the continuum limit, since only a finite number of perturbative graphs need be canceled.

If we periodically identify one of the Poincar\'e directions of the $D=4$ theory, we would expect its low energy physics to be given by the $D=3$ theory.  Correspondingly on the gravity side, the periodic dimension pinches off near the horizon due to the $L^2/z^2$ in the metric, and a $T$-duality brings us to the D2-brane geometry.  One can repeat this to get down to D0-branes, and I think that it is accurate to say that the BFSS Matrix Theory conjecture~\cite{Banks:1996vh} is implied by AdS/CFT duality.  It is less clear that the reverse is true, because the BFSS duality focuses on certain extreme low energy observables.

If we move up in dimension to D4-branes, the order of things is reversed.  The gauge theory is classical at low energy, while at higher energies the effective description is the D4-brane metric, and at higher energies still the dilaton becomes large and we go over to the M5-brane geometry.  We do not need the duality to solve the low energy physics because it is free, but it provides a UV completion.  Higher dimensional branes become more intricate.

Question: How can the $N^2$ D2-brane entropy go over to the $N^{3/2}$ M2-brane entropy?  Ref.~\cite{Itzhaki:1998dd} finds that the black 2-brane entropy is
\begin{equation}
s_{\rm black\,D2} \sim  N^2 T^2 (\lambda/T)^{-1/3} \,.
\end{equation}
Unlike the D3-brane, this depends on the 't Hooft coupling, and as $\lambda$ increases more degrees of freedom freeze out.  Why the D2 should be different from the D3 in this regard is a puzzle!   But because $\lambda$ is dimensional, this implies an extra $T^{1/3}$ dependence.  Exercise: by following through the calculations in Ref.~\cite{Itzhaki:1998dd}, show that this  suppression at low temperatures is just enough to make the D2 result match on to $N^{3/2}$ at the crossover temperature between the D2 and M2 pictures.

\subsection{D2-D6}

(The discussion in this and the following section is based in part on unpublished work with E. Silverstein.)  

For a $D=3$ gauge theory with matter, the one-loop running is
\begin{equation}
\mu\partial_\mu \gamma^2 = -\gamma^2 + b_0 \gamma^4 \,, \label{d31loop}
\end{equation}
where $\gamma^2 = g_{\rm YM}^2 / \mu$ is the dimensionless gauge coupling.  The constant $b_0$ is proportional to the one-loop $\beta$-function from $D=4$.  If $b_0 > 0$, meaning the that $D=4$ theory is free in the IR, then the one-loop equation~(\ref{d31loop}) exhibits an IR fixed point at nonzero coupling $\gamma^2 = 1/b_0$.  This is not reliable in general, but if we take an $N$-vector of matter fields the one-loop term dominates higher orders and we can believe the result.  Thus we generate a large class of conformal theories in $D=3$~\cite{Appelquist:1989tc}, gauge theory analogs of the Wilson-Fisher fixed point.

We can build such theories with branes by taking D2-branes lying within D6-branes, i.e. they are extended in the directions
\begin{equation}
\begin{array}{ccccccccccc}
&0&1&2&3&4&5&6&7&8&9 \\
\mbox{D2:}&\times&\times&\times&&&&&&& \\
\mbox{D6:}&\times&\times&\times&\times&\times&\times&\times&&&
\end{array} 
\end{equation}
This preserves half the supersymmetry of the D2's, namely $D=3,\ {\cal N} = 4$.  The 2-2 strings are the $SU(N_2)$ reduction of the $D=4$ theory and give $b_0 = 0$ for this group.  The 2-6 strings provide $N_6$ fundamentals and give a positive contribution to $b_0$.  The 6-6 strings are free at low energy, because of the high dimension of the D6-brane, and do not of course contribute to the $SU(N_2)$ running.  Thus, for $N_6 \gg N_2$ there is a weakly coupled fixed point.  In general one cannot guarantee that this fixed point survives to strong coupling, but here there is enough supersymmetry to assure that it does.  Moreover there is a gravity dual~\cite{Pelc:1999ms}.  Normally it is difficult to describe such localized branes explicitly, but here the M-theory lift of the D6-brane allows to construct the solution as the ${\mathbb Z}_{N_6}$ orbifold of the M2-geometry of charge $N_2 N_6$~\cite{Itzhaki:1998uz}.  For $N_6 \ll N_2 \ll N_6^5$ there is a weakly coupled IIA description, and for $N_6^5 \ll N_2$ the M theory picture is the valid one.  These same scalings are found in a different orbifold of this geometry studied in Ref.~\cite{Aharony:2008ug}.

\subsection{D2-D8}

Now consider D2-D8,
\begin{equation}
\begin{array}{ccccccccccc}
&0&1&2&3&4&5&6&7&8&9 \\
\mbox{D2:}&\times&\times&\times&&&&&&& \\
\mbox{D8:}&\times&\times&\times&\times&\times&\times&\times&\times&\times&
\end{array} \,.
\end{equation}
This case is similar to the preceding, except that there is no supersymmetry, and the 2-8 strings are all fermionic.  This sounds a good starting point for a Fermi liquid: turn on a chemical potential for these strings and the geometry will tell us what happens when they are in interaction with a strong gauge coupling (in the D2-D6 case, there are also 2-6 bosons, and energetically it is likely that the charge will be carried predominantly be these).

An effective potential analysis along the lines of that in Sec.~2.1 indicates that there is the possibility of an $AdS_4$ solution before we turn on the chemical potential.  The analysis is crude, though, because there is less symmetry (the D8's lie on a particular $S^5$ within the $S^6$ surrounding the D2's).

Further, there are several instabilities that may keep us from getting to the desired strong coupling phase.  The system may be unstable along the Coulomb branch direction, the D2-branes separating (E. Silverstein has christened this seasickness because the Fermi sea make the AdS throat eject the branes).  The D8's may be repelled from the D2's, which is chiral symmetry breaking, a long-standing issue in carrying these $D=3$ gauge fixed points to strong coupling.  Shape modes of the $S^6$ surrounding the D2's may become BF-forbidden tachyons; this is a common occurence in nonsupersymmetric AdS spaces~\cite{DeWolfe:2001nz}.  These instabilities may be present even before the chemical potential is turned on, and the fermion density seems to make some worse (e.g. the seasickness along the Coulomb branch).

It seems likely that one can eventually, by adding additional ingredients, stabilize any of these and get to a Fermi liquid dual, at least one that is metastable (one additional knob is to add D6-branes in various orientations).  Therefore let us ignore them as far as we can and turn on a chemical potential.  Something nice then happens~\cite{Kulaxizi:2008jx}, in that the system does exhibit at least one feature of a Fermi surface: there are particle-hole states of finite momentum and arbitrarily small energy, as one gets by taking a fermion from just below the surface and moving it to a point just above the surface somewhere else.  The calculation is this: turning on a chemical potential introduces a radial electric field on the D6-brane.  One measures the correlator of the fermion current (dual to a 6-6 string) by expanding the D-brane Dirac-Born-Infeld action, and because of the special form of this action (which in particular has a maximum electric field) the correlator has the distinctive behavior of an imaginary part that goes down to zero frequency at finite momentum.

There are still several puzzles, however: 1) this behavior is also seen in the D2-D6 system, where a Fermi liquid might not have been expected; 2) in Fermi liquids there is a maximum momentum of $2k_{\rm F}$, the maximum distance an electron can be moved, while here there is no maximum --- perhaps this is a strong-coupling effect; 3) the large field near the origin backreacts on the metric in a divergent way, and the resolution of this singularity is not understood in general, it could gap the surface; 4) even when the backreaction is parametrically suppressed by taking $N_2$ large, the DBI action breaks down near the horizon due to the field gradients.

Question: What about $T$-dual configurations, like D3-D7 in the nonsupersymmetric orientation
\begin{equation}
\begin{array}{ccccccccccc}
&0&1&2&3&4&5&6&7&8&9 \\
\mbox{D3:}&\times&\times&\times&\times&&&&&& \\
\mbox{D7:}&\times&\times&\times&&\times&\times&\times&\times&\times&
\end{array}  \,?
\end{equation}
Good question: the intersection is still $D=3$, and the fermions live there, but the gauge fields now move in $D=4$.  Things are less IR singular but still interesting: the perturbative analysis seems more tractable.  So maybe this is another interesting case.  For D4-D6,
\begin{equation}
\begin{array}{ccccccccccc}
&0&1&2&3&4&5&6&7&8&9 \\
\mbox{D4:}&\times&\times&\times&\times&\times&&&&& \\
\mbox{D8:}&\times&\times&\times&&&\times&\times&\times&\times&
\end{array}  \,,
\end{equation}
the gauge fields move in $D=5$ and are classical at low energy, they do not give interesting dynamics.

\subsection{Fractional D3-D7}

Finally I mention a case that addresses the generality of AdS/CFT.  If we replace the ${\cal N}=4$ gauge theory with ${\cal N}=2$, and add $N_f = 2N_c$ fundamentals, we get another theory which is conformal for all values of the coupling.  We can get this from $N_c$ fractional D3-branes at an orbifold point plus $N_f$ D7-branes.  However, the conformal limit is singular and suggests a $T$-duality to a IIA configuration~\cite{Grana:2001xn}.  Even then, an effective potential analysis suggests that the scales of both the compact space and the AdS space are stringy, meaning that the anomalous dimensions remain of order one even at large coupling, they do not grow as a power of the coupling.  Thus, it appears that the necessary condition of large operator dimensions is rather special.  Further this seems to go against the principle that whenever a coupling becomes very large we can find a small expansion parameter in a different picture.  Ref.~\cite{Gadde:2009dj} discusses this system in much more detail, and proposes a more intricate dual picture.


\section*{Acknowledgments}
I would like to thank Allan Adams, Alex Buchel, Tom Faulkner, Sean Hartnoll, Idse Heemskerk, Gary Horowitz, Jacopo Orgera, Amanda Peet, Joao Penedones, Eva Silverstein, Matt Strassler and James Sully for collaborations on various aspects of this subject, and Tom Banks, Gary Horowitz, Hong Liu, Don Marolf, Dam Son, and Sho Yaida for useful discussions.  I would also like to thank the students at TASI for their incessant questions and all-around enthusiasm, which made this lecture series a rewarding experience.  This work was supported in part by NSF grants PHY05-51164 and PHY07-57035.

\end{document}